\documentstyle[aps,prl,twocolumn,epsf]{revtex}
\begin{document}
\input{psfig.sty}
\draft
\newcommand{\lscoo}{\mbox{${\rm La_{1.5}Sr_{0.5}CoO_4}$}}
\newcommand{\cotwo}{Co$^{2+}$}
\newcommand{\cotri}{Co$^{3+}$}
\newcommand{\bQ}{\mbox{${\bf Q}$}}
\newcommand{\h}{\mbox{$\frac{1}{2}$}}
\newcommand{\th}{\mbox{$\frac{3}{2}$}}
\def\vtau{\mbox{\boldmath$\tau$}}

\twocolumn[\hsize\textwidth\columnwidth\hsize\csname
@twocolumnfalse\endcsname

\title{Independent freezing of charge and spin dynamics in
La$_{1.5}$Sr$_{0.5}$CoO$_4$. }

\author{I.~A.~Zaliznyak$^{1}$, J.~P.~Hill$^{1}$, J.~M.~Tranquada$^{1}$,
 R.~Erwin$^{2}$, Y.~Moritomo$^3$}
\address{
 $^1$Department of Physics, Brookhaven National Laboratory, Upton,
 New York 11973-5000 \\
 $^2$National Institute of Standards and Technology, Gaithersburg,
 Maryland 20899\\
 $^3$Center for Integrated Research in Science and Engineering (CIRSE)
 and Department of Applied Physics, Nagoya University, Nagoya
 464-01, Japan
 }

\date{\today}
\maketitle

\begin{abstract}

We present elastic and quasielastic neutron scattering
measurements characterizing peculiar short-range charge-orbital
and spin order in the layered perovskite material \lscoo. We find
that below $T_c\approx 750$~K holes introduced by Sr doping lose
mobility and enter a statically ordered {\it charge glass} phase
with loosely correlated checkerboard arrangement of empty and
occupied $d_{3z^2-r^2}$ orbitals (Co$^{3+}$ and Co$^{2+}$). The
dynamics of the resultant mixed spin system is governed by the
anisotropic nature of the crystal-field Hamiltonian and the
peculiar exchange pattern produced by the orbital order. It
undergoes a {\it spin freezing} transition at much a lower
temperature, $T_s\lesssim30$~K.

\end{abstract}

  %(Antiferromagnetics). }

]

The instability of the doped transition-metal oxides towards
formation of cooperative charge and spin ordered phases is one of
the central phenomena in the physics of colossal magnetoresistance
(CMR) materials and high temperature superconductors. Charge
segregation into lines \cite{TranquadaWakimotoLee} which separate
stripes of antiphase antiferromagnetic domains in
La$_{2-x}$Sr$_x$CuO$_{4+y}$ (LSCO) cuprates at small $x$, is,
according to some theories, key to their superconductivity
\cite{Kivelson98Zaanen99}. In perovskite manganates, built of
essentially isostructural Mn-O layers \cite{Schiffer95}, doping
first destroys the cooperative Jahn-Teller (JT) distortion of the
orbitally-ordered insulating state \cite{Murakami98}, and induces
a transition to a ferromagnetic metal (FM) \cite{Millis}. The
extremely strong response of transport properties in the FM phase
to an applied magnetic field, termed CMR, results from the strong
Hund's coupling of charge carriers to the Mn$^{4+}$ core spins
\cite{ZenerAnderson}. Then, at doping $x\approx 0.5$ an
instability towards another kind of charge-orbital order, with a
checkerboard arrangement of Mn$^{3+}$/Mn$^{4+}$ ions, results in
an antiferromagnetic (AFM) insulating ground state. This doping
level is argued to be of greatest technological relevance
\cite{Schiffer95}, since at half-doping transport properties
usually show the strongest response to a magnetic field. To
understand the physics of charge/spin ordered (CO/SO) phases, and,
in particular, to answer the question of whether CO/SO are
independent instabilities or closely coupled, the relative
importance of the various interactions (superexchage, double
exchange, Coulomb repulsion, JT distortion) needs to be clarified.
Experimentally this can be done by studying different compounds in
which the strengths of the above interactions vary.

The first complete characterization of the spin and charge order
in the half-doped regime by means of neutron diffraction was
reported for La$_{1.5}$Sr$_{0.5}$MnO$_{4}$ \cite{Sternlieb96}.
Nuclear scattering accompanying the CO was found below
T$_{co}\approx 217$~K, where the steep rise in resistivity
manifests the transition to an insulating state. A magnetic signal
consistent with simple collinear antiferromagnetic (AFM) order
appeared below T$_{so}\approx110$~K. CO was confirmed in a recent
X-ray resonant scattering experiment \cite{Murakami98a}, which has
also provided the first direct evidence of the concomitant orbital
order (OO). Similar charge, orbital and magnetic order was also
observed in the ground state of half-doped pseudocubic manganates
\cite{ZimmermanJirak}. In most cases, the charge orders at a
somewhat higher temperature than the spins do, but whether the CO
instability occurs independently or is driven by the
magnetic/orbital fluctuations is a topic of continuing debate
\cite{SolovyevBrink,vanDuin98}.

Here we report a detailed neutron scattering study of the closely
related layered cobalt oxide La$_{1.5}$Sr$_{0.5}$CoO$_{4}$. We
argue that strong single-ion anisotropy, mediated by the
relativistic spin-orbit coupling usually neglected in cuprates and
manganates, effectively decouples charge ordering from low-energy
spin fluctuations. This is reflected in the spectacular difference
in CO and SO transition temperatures, $T_{co}/T_{so}\gtrsim 20$ in
this material. In fact, strong planar anisotropy leads to a {\it
quenching of the spin angular momentum} on the Co$^{3+}$ sites, in
full analogy with the well-known quenching of the orbital momentum
by the crystal field. Combined with the checkerboard arrangement
of the doped holes in the CoO$_2$, planes this makes the spin
system of La$_{1.5}$Sr$_{0.5}$CoO$_{4}$ a strongly frustrated
square lattice antiferromagnet.

Electronic configurations of Co$^{2+}$ ($3d^7$) and Co$^{3+}$
($3d^6$) ions are related to those of Mn$^{4+}$ and Mn$^{3+}$,
respectively, by virtue of electron-hole symmetry. However, in the
case of cobalt, Hund's rule is in close competition with the cubic
crystal field, the latter splitting the $3d$ level into a
lower-lying $t_{2g}$ triplet and an $e_{g}$ doublet of
$d_{x^2-y^2}$ and $d_{3z^2-r^2}$ orbitals. As a result, even
though Co$^{2+}$ is in the high-spin state ($t_{2g}^5e_g^2$,
S=3/2), Co$^{3+}$ obtained by adding the fourth hole may be either
in the high $t_{2g}^4e_g^2$ (HS, S=2), intermediate
$t_{2g}^5e_g^1$ (IS, S=1), or low spin state $t_{2g}^6$ (LS, S=0)
\cite{Moritomo98}. Transitions between different spin states are
not uncommon in Co compounds. If LS is the ground state, a
decrease in free energy due to the higher magnetic entropy may
drive transitions to IS and HS states with increasing temperature,
as observed in LaCoO$_3$ \cite{Asai98}. For the purpose of the
present study, however, the spin state of the Co$^{3+}$ ions is
not important at sufficiently low temperatures where spin order
occurs. Indeed, {\it any integer spin} will be {\it frozen in a
singlet state} by strong planar anisotropy, rendering ions
effectively non-magnetic for low-energy fluctuations
\cite{Lindgard93}.

We studied a piece (0.48 g) of high quality single crystal of
\lscoo\ grown by the floating-zone method \cite{Moritomo98}. In
the temperature range $6$ K$\lesssim T\lesssim 600$ K covered in
our experiment, the crystal remained in the tetragonal ``HTT" phase
(space group $I4/mmm$), with low $T$ lattice parameters $a=3.83$~\AA\ and
$c=12.5$~\AA. However, to index the superlattice peaks it is convenient to
choose a unit cell that is twice as large
($\sqrt{2}a\times\sqrt{2}a\times c$) corresponding to space group $F4/mmm$.

Most of the experiments were performed on the BT2 and BT4 3-axis
thermal neutron spectrometers at the NIST Center for Neutron
Research. Some preliminary measurements were done on the H8
spectrometer at the High Flux Beam Reactor at Brookhaven. In all
cases, PG(002) reflections were used at the monochromator and
analyser, supplemented by PG filters to suppress the higher-order
contamination. The energy of the scattered neutrons was fixed at
$E_f=14.7$ meV. Beam collimations before the monochromator and
after the analyser were kept at $\approx 60'$ and $\approx 100'$,
respectively, while around the sample they were set either to
$20'-20'$ or to $42'-62'$ depending on the resolution required.
Our sample is a cylinder with axis parallel to [010] direction,
$D\approx 4$ mm, $L\approx 3$ mm. It was mounted in the displex
refrigerator with axis vertical, allowing wavevector transfers in
the $(h0l)$ reciprocal lattice plane. Rocking curves about the
[010] direction reveal a mosaic broadening of less than
$0.25^\circ$. Normalization of the scattered intensity was done
using the incoherent scattering from a vanadium sample.

Figure 1 shows a survey of the low-temperature elastic scattering
in \lscoo. Peaks of magnetic and structural origin are easily
distinguished by their characteristic wavevector dependences. The
magnetic intensity is proportional to the Fourier-transform of the
density of unpaired electrons (the magnetic form factor) squared,
$|f(Q)|^2$, which becomes weaker at large $Q$, while the intensity
from small nuclear displacements increases as $\sim Q^2$. This
leads us to conclude that the peaks in Fig.~1(b), centered at
slightly incommensurate positions $\vtau\pm {\bf Q}_m$, ${\bf Q}_m
= (0.5+\epsilon,0,1)$ with $\epsilon\approx 0.017$, are magnetic,
while stronger feature in Fig.~1(c) and features in Fig.~1(d)
result from the modulations of the atomic positions accompanying
the charge order of \cotwo/\cotri\ ions. None of the
superstructure peaks are resolution-limited in $Q$, indicating
finite-range {\it correlated glass} type modulations. They are
static on the time scale $\delta t \lesssim 1$ ps as determined by
the energy resolution of our experiment. The diffuse nature of the
scattering is most apparent in scans along the ${\bf c}^*$ ([001])
direction, which reflects the anisotropy of the correlation
lengths: in-plane correlations are much better developed than
inter-plane (see Table~1).

%==============================Fig.1==================================
\begin{figure}[t] \noindent\vspace{0.1in}
\parbox[b]{3.4in}{\psfig{file=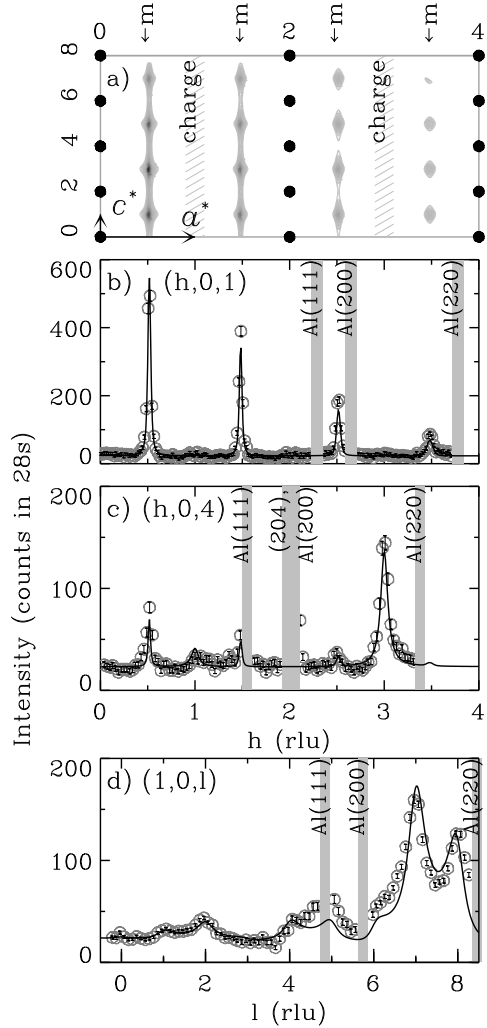,height=4.75in}
\vspace{0.45in} \noindent \caption{Survey of the elastic
scattering in the $(h0l)$ reciprocal plane of \lscoo. (a) Solid
circles show allowed nuclear Bragg reflections, hatched areas mark
regions where additional nuclear scattering due to the charge
order is observed. The grey-scale map represents calculated
magnetic intensity. (b)-(d) representative scans through the peaks
of magnetic and charge-orbital scattering collected at $T=10$~K
(b) and $T=6.1$~K (c),(d). Points in grey-shaded regions
contaminated with aluminum scattering are removed. Solid curves
show the fits discussed in text.}}
\end{figure} \vspace{-0.1in}
%=====================================================================

The incommensurability of the charge-order peaks, if any, is too
small to be directly resolved, because of the short correlation
length and the overlap of peaks at $\vtau\pm {\bf Q}_c$. However,
the scattering is consistent with the modulation wave vectors
${\bf Q}_c = (1+2\epsilon,0,l)$ and $(0,1+2\epsilon,l)$, with
$l=0$ or 1. If $\epsilon$ is zero, the charge order corresponds to
a checkerboard superstructure within each CoO$_2$ plane, as
indicated in Fig.~2(c).  The magnetic modulation corresponds to
almost antiferromagnetic order on either the \cotwo\ or \cotri\
sublattice, with a superposed long-wavelength modulation along
{\bf a} (or {\bf b}).

%==============================Fig.2==================================
\begin{figure}[t] \noindent
\parbox[b]{3.4in}{\psfig{file=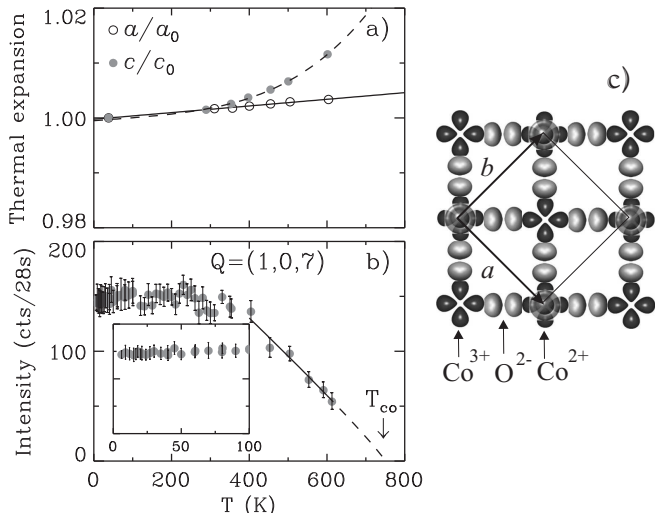,width=3.4in}
\vspace{0.in} \caption{(a) Relative change in the in-plane (open)
and inter-plane (closed symbols) lattice spacings refined in the
longitudinal scans through (200) and (006) Bragg reflections. (b)
Temperature dependence of the intensity of the diffuse peak at
$\bQ=(107)$ associated with the checkerboard charge-orbital order.
Insert expands the low-$T$ region, $T\leq 100$~K. (c) Schematic
drawing of Co-O bonding orbitals and checkerboard order of
\cotwo/\cotri\ valence in $a-b$ plane.}}
\end{figure} \vspace{-0.1in}
%=====================================================================

For the quantitative analysis of nuclear scattering accompanying
the charge order, we use the following correlation function
between the displacement $\mbox{\boldmath$\varepsilon$}_{\mu{\bf
r}}$ of atom $\mu$ at the lattice point ${\bf r} =r_a\hat{\bf a} +
r_b\hat{\bf b} + r_c\hat{\bf c}$ and
$\mbox{\boldmath$\varepsilon$}_{\mu'{\bf 0}}$ of atom $\mu'$ at
${\bf 0}$:
\begin{equation}
  \langle \varepsilon_{\mu{\bf r}}^\alpha\varepsilon_{\mu '{\bf 0}}^\beta
  \rangle =\varepsilon_\mu^\alpha \varepsilon_{\mu '}^\beta
  \cos({\bf Q}_c{\bf r}) \;
  e^{-(|r_a|+|r_b|)/\xi_\|^{co}-|r_c|/\xi_\bot^{co}},
  \label{chargeorder}
\end{equation}
describing a short-range harmonic modulation. Brackets denote an
average over the crystal volume; $\alpha,\beta = x,y,z$; and
$\xi_\|^{co}$ and $\xi_\bot^{co}$ are the charge-order correlation
lengths parallel and perpendicular to the planes, respectively.
The scattering cross-section obtained from Eq.~(\ref{chargeorder})
upon appropriate Fourier transformation has the form of factorized
``lattice Lorentzians" \cite{Fisher} in three directions, weighted
with $\left|\sum_\mu ({\bf
Q}\cdot\mbox{\boldmath$\varepsilon$}_\mu) b_\mu e^{-i{\bf
Q}\cdot{r}_\mu}\right|^2$, similar to that used to describe
finite-range stripe correlations in LSCO
\cite{TranquadaWakimotoLee}. Here $b_\mu$ is the scattering length
(including the Debye-Waller factor) of the nucleus at position
${\bf r}_\mu$ in the unit cell.  The $Q$-depedence of the
structural diffuse intensity evident in scans along ${\bf c}^*$,
Fig~1(d), suggests that it arises from breathing-type distortions
of the oxygen octahedra surrounding the Co ions. Indeed, it
appears sufficient to consider displacements of the in-plane
oxygens O(1), $\varepsilon^{x,y}_{O(1)}$, and apical oxygens O(2),
$\varepsilon^{z}_{O(2)}$, along the corresponding Co-O bonds, to
obtain a good description of the observed ``charge" scattering.
Solid curves in Fig.~1(c),(d) present the neutron intensity
calculated from Eq.~(\ref{chargeorder}) after appropriate
normalization and correction for the spectrometer resolution. The
parameters $\varepsilon^{x,y}_{O(1)} = 0.042(4)$ \AA,
$\varepsilon^{z}_{O(2)} = -0.066(2)$ \AA, and the correlation
lengths shown in Table~1 were obtained in a global fit of all
scans around $\vtau\pm {\bf Q}_c$. The error bars shown do not
include any possible systematic error from the vanadium
normalization, which relied upon a knowledge of the vertical beam
divergences (note the absence of arbitrary scaling factors between
calculated and measured intensities in all panels of Fig.~1).

Figure~2 shows the temperature dependence of the peak intensity of
the diffuse ``charge" scattering and lattice thermal expansion in
the {\bf a} and {\bf c} directions. Although we could not reach
the CO melting temperature, an estimate of $T_{co}\approx 750$ K
is obtained by extrapolation shown in Fig.~2(b). Strong nonlinear
decrease of the $c$ lattice spacing which accompanies the charge
order is indirect evidence of the concomitant {\it orbital order}.
This is because the checkerboard ordering of empty and occupied
out-of-plane $d_{3z^2-r^2}$ orbitals, combined with the
body-centered stacking of Co-O planes,
\mbox{\boldmath$\delta$}$=[\h0\h]$, allows a reduction of the
inter-plane spacing.

We analyze the magnetic scattering cross section starting from a
spin-spin correlation function similar to Eq.~(\ref{chargeorder}).
By simultaneously fitting all of the magnetic peaks measured at
$T=10(3)$~K, we obtain a SO incommensurability $\epsilon =
0.017(1)$ and the correlation lengths shown in Table 1. We also
find that the spin structure is planar, and isotropic within the
$a-b$ plane. This agrees with the static susceptibility data
\cite{Moritomo98}, which yield an estimated $D\gtrsim 400$ K for
the XY-type anisotropy energy. For $T\ll D$ this restricts any
integer spin to a singlet state \cite{Lindgard93}. Further support
for the conclusion that \cotri\ ions are effectively non-magnetic
is provided by small value of the frozen magnetic moment,
$\langle\mu\rangle = 1.4(1)\mu_B$ per Co site refined for a single
domain model. Assuming two equivalent ${\bf Q}_m$ domains with
half of the sites occupied by magnetic \cotwo\ ions, we obtain
$\mu_{\rm Co^{2+}} \approx 2.9\mu_B$, within 20\% of what is
expected for S=3/2.

The main role of the "non-magnetic" \cotri\ ions is to bridge the
\cotwo\ ions, providing effective antiferromagnetic coupling. As
shown in Fig.~2(c), there are two \cotwo--O--\cotri--O--\cotwo\
exchange pathways between nearest neighbor (nn) \cotwo\ ions, and
one between next-nearest neighbors (nnn). As a result, the spin
system appears to be a quasi-two-dimensional square-lattice
antiferromagnet with nnn to nn exchange ratio $J_2/J_1\approx
0.5$; {\it i.e.}, it is in the critical region where frustration
destroys

\begin{table}[t]
\label{table1} \caption{Anisotropic correlation lengths of the
frozen magnetic and charge-orbital order and approximate positions
of the corresponding peaks in reciprocal space.}
\begin{tabular}{lccc}
 & $\xi_\|$ (\AA) & $\xi_\bot$ (\AA) & Peak position \\
\tableline
 charge & 28(2) & 8.1(7) & $\approx(2h\pm 1,0,l)$ \\
 magnetic & 79(3) & 10.7(3) & $\approx(2h\pm 0.5,0,2l+1)$ \\
\end{tabular}
\end{table}

\noindent N\'{e}el order \cite{Zhitomirsky96}. This provides a
natural explanation for the short spin-spin correlation length in
the $a-b$ plane and the peculiar {\it spin freezing} transition
revealed by the temperature dependences shown in Fig.~3. Despite a
somewhat ``order parameter"-like dependence of the (quasi)elastic
magnetic peak intensity, the correlation lengths do not diverge.
In fact, the defining feature of the transition is the
disappearance of the {\it energy width} $\Gamma_E$ of the magnetic
scattering, {\it i.e.} a divergence of the relaxation time for the
spin fluctuations [Fig. 3(b)]. Upon appropriate correction for the
instrumental resolution, we are able to refine the $\Gamma_E(T)$
dependence down to $\sim 0.05$ meV. From the power-law fit
$\Gamma_E(T)\sim (T-T_{so})^\zeta$ shown in Fig.~3(b) we estimate
$\zeta=2.7(3)$ and $T_{so}\approx 30$ K.

%============================Fig.3====================================
\begin{figure}[t] \noindent
\parbox[b]{3.4in}{\psfig{file=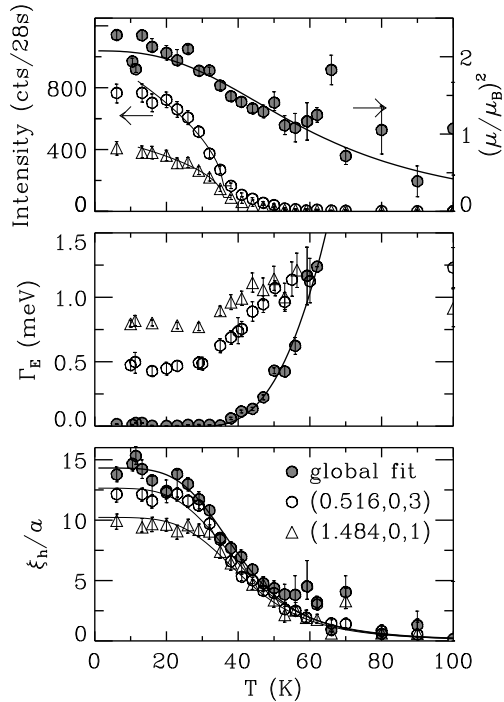,height=4.5in}
\vspace{-0.67in} \caption{Temperature dependencies of the
parameters of magnetic scattering. Open symbols result from simple
fits of the raw data with Lorentzian (a),(c) and Gaussian (b)
curves. Shaded circles were obtained from the resolution-corrected
global fit of all scans discussed in the text. (a) peak intensity
(left scale) and squared magnetic moment per Co site (right
scale), (b) energy width, and (c) the in-plane correlation length.
Lines in (a),(c) are guides for the eye.}}
\end{figure} \vspace{-0.1in}
%=====================================================================

Finally, to underscore the significance of these results we
compare the charge and spin order of \lscoo\ with that of
La$_{1.5}$Sr$_{0.5}$MnO$_4$ and other manganates. Firstly, the
charge order in \lscoo\ occurs independently of the magnetic
order, arising at temperatures $\gtrsim 20$ times higher than the
characteristic energy scale of the cooperative spin fluctuations.
Secondly, compared to the CO in the manganate, it has a much
shorter correlation range, $\xi_{co}({\rm Mn}) / \xi_{co}({\rm
Co}) \gtrsim 10$, but, surprisingly, results in an even stronger
charge localization with an activation behavior of electrical
conductivity with $E_a\sim 6000$ K \cite{Moritomo98}. Thirdly, as
a result of the checkerboard order of \cotwo/\cotri\ ions and the
strong $XY$ anisotropy, the spin system is a stack of weakly
interacting square antiferromagnets with nearly critical
frustration, $J_2/J_1\approx 0.5$. In contrast to
La$_{1.5}$Sr$_{0.5}$MnO$_4$, where simple collinear AFM order
occurs, the cobaltate undergoes a freezing transition into a
short-range correlated, incommensurate spin-glass state. In fact,
this is a rare example where the evolution of the critical spin
fluctuations in the course of glassification can be studied by
neutron scattering over a dynamic range of about two decades. We
intend to further investigate the dynamics of the spin freezing
transition in the near future.

We thank NIST Center for neutron Research for hospitality during
the experiments. This work was carried out under Contract NO.
DE-AC02-98CH10886, Division of materials Sciences, US Department
of Energy.

\end{document}